\documentclass[9pt,twocolumn,twoside]{osajnl}

\DeclareMathOperator\erf{erf}
\DeclareMathOperator\erfc{erfc}

\journal{ao} 

\setboolean{shortarticle}{false}

\title{Impact of spectral noise shape and correlations of laser beam jitter on acquiring optical links in space}
\author[1]{Gerald Hechenblaikner}
\affil[1]{Airbus Defence and Space, Claude-Dornier-Strasse, 88090 Immenstaad, Germany; Gerald.Hechenblaikner [at] airbus.com}

\begin{abstract}
We investigate how the probability of acquiring an optical link between a scanning and a target spacecraft depends on the spectral shape, power and dimensionality of the beam jitter, as well as on the choice of detector integration time, beam detection radius and scan speed. 
For slow scans and long integration times, the probability of failure (Pfail) is determined by the integrated jitter power up to a critical frequency, which we verify by comparing the results of an analytical model to those of Monte Carlo simulations. Jitter above the critical frequency leads to a loss of correlation between integration windows and decreases Pfail for both, 1d (radial) and 2d (radial and tangential) jitter, as long as the RMS jitter amplitude does not exceed the beam diameter.  In the opposite limit of fast scans and short integration times, emergent correlations between jitter fluctuations on two adjacent scanning tracks also decrease Pfail. The analytical model is additionally used to assess the effect of multiple overlapping tracks and the impact of target drifts in the uncertainty plane.
\end{abstract}

\setboolean{displaycopyright}{false}

\begin{document}

\maketitle

\section{Introduction}
\label{sec:intro}
The number of optical links established between satellites in space has rapidly increased over the past two decades, from first communication experiments between the Artemis and SPOT4 satellites via the the SILEX system \cite{fletcher1991silex} in 2001, 
to current routine high-performance links between the Alphasat satellite on geostationary-orbit (GEO) and the Sentinel-1A/1B satellites on low-earth orbit (LEO)\cite{benzi2016optical}. The future European Data Relay System comprises a network of satellites equipped with optical terminals that provide global coverage for fast data transmission between network nodes in order to support high resolution imaging for crisis management and security applications, offering secure point-to-point communication over large distances\cite{hauschildt2019global}.
Additionally, optical links are used in missions to accurately measure the earth gravitational field, such as GRACE FO\cite{heinzel2017laser} or the planned NGGM\cite{nicklaus2020laser}, as well as in the future gravitational wave observatories in space, such as TAIJI \cite{luo2020taiji} or LISA \cite{danzmann2000lisa}.
All these missions require the optical links to be acquired before they can be used for either communication or scientific measurements.

\subsection{The acquisition process}
In the most commonly used ``beaconless acquisition'' architecture, the same optical beam that is used for communication is also used for acquisition\cite{hindman2004}, which greatly reduces the required resources compared to using a dedicated search beacon. However, due to the narrow angular beam width it is necessary to scan the uncertainty area using a specific search pattern, making this search method more complex. This paper focuses on critical aspects in relation to beaconless acquisition. 

In a typical acquisition scenario, one of the two spacecraft (SC) is scanning the uncertainty region in angular space, referred to as ``uncertainty cone'', where the receiving SC is expected to be. If the scanning beam is incident on the receiving SC within its acquisition sensor field-of-view, the receiving SC detects it and re-orients itself towards the direction of the received light signal. The required pointing accuracy is achieved if the detector resolution and the signal-to-noise ratio of the detected signal are sufficiently large. In the next step the scanning SC likewise detects the beam from the receiving SC, stops its scan, and re-orients itself towards the receiving SC. This completes the ``spatial acquisition'' phase which is typically followed by the conceptually simpler ``frequency acquisition'' phase, where one SC locks its laser to the beam received from the other SC.

\subsection{Previous models and scope of this paper}
First simulations of acquisition search patterns, including raster and spiral scans, were made in \cite{scheinfeild2000}, where the impact of noise was also analyzed in relation to distortion of the scanning path. Requirements for beaconless acquisition were analyzed in \cite{hindman2004}, in particular with respect to the divergence angle of the scanning beam.  
The mean search time was computed with an analytical model in \cite{li2011analytical}, assuming a series of repetitive scans is required in order to increase the detection probability.
This model was recently extended to include the impact of vibrations on the mean search time in \cite{ma2021satellite}.
The probability of ``hitting'' the receiving SC at a specific location with the vibrating scanning beam was studied in \cite{friederichs2016vibration}, but the more relevant probability average over the SC uncertainty distribution was not further analyzed. 
In our  recent paper \cite{hechenblaikner2021analysis} we developed an analytical model for the overall detection probability in relation to the laser beam width, the scanning spiral pitch, and the RMS fluctuations of the laser beam.\\
In this paper we analyze critically important effects and relationships between acquisition parameters that -to our best knowledge- have never been studied before, focusing on the dependency of the detection probability on correlations of the beam jitter which may either occur between spiral tracks or between individual integration windows.
Previous studies generally made the approximation that the scanning beam is detected if the receiving SC is located within a given radius from the beam center at any point during the acquisition scan, which represents the limit of infinitely short integration times. Model predictions for this case generally differ from the actual physical case of finite integration times that are required in order to accumulate sufficient energy for detection which we investigate in this paper. Furthermore, our analyses are based on band-limited noise of different spectral shapes, which also leads to quite different results, in particular for slow scan speed and long integration times, compared to unlimited band-width considered in previous studies\cite{friederichs2016vibration, ma2021satellite}.\\
We first extend the previous model\cite{hechenblaikner2021analysis} to account for beam overlaps across multiple tracks, allowing us to analyze the effect of drifts of either scanning beam or receiving SC in the uncertainty plane, which we find to have a significant impact on the probability of detection.
Performing comprehensive Monte Carlo simulations, we then investigate the impact of the spectral properties of beam jitter on the detection probability, in particular the noise shape and its associated auto-correlation function. We find that the analytical model can be applied for noise shapes in between two limiting cases for which we derive parametric scaling equations.\\
In one extreme, if the width of the auto-correlation function is larger than the time of one full spiral revolution, this correlates fluctuations between two adjacent tracks and significantly increases the probability of detection. This condition is the typical operational regime of optical communication terminals with a fast beam-steering mirror, large beam intensities in the far field, and correspondingly short integration times\cite{sterr2011beaconless, friederichs2016vibration}.\\
In the other extreme, if the width of the auto-correlation is smaller than the time it takes for the beam to sweep past the location of the receiving SC, the integration windows within this period become uncorrelated, which also increases the detection probability. This regime may be entered by science missions with low beam powers in the far-field, requiring large integration times and slow scan speeds\cite{cirillo2009control,heinzel2017laser}. In addition, we analyze the different impact of 1d (radial) and 2d (radial and tangential) jitter, as well as the effect of the nonlinear intensity distribution of the laser beam profile, on the probability of detection.\\
The major aim of this paper is to enable the system architect to select optimal acquisition parameters across a wide range of different possibilities by considering the parametric relationships and scaling laws discussed in the following sections.

\section{The Radius of detection}
\label{sec:R_d}
\begin{figure}
\centerline{\includegraphics[width=0.8\columnwidth]{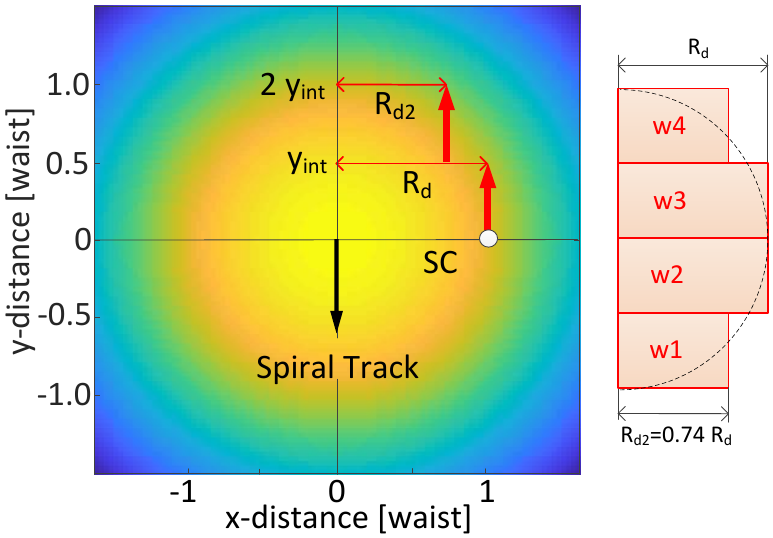}}
\caption  {The laser beam intensity profile in the uncertainty plane. When the laser beam sweeps over the location of the receiving SC (white circle), the detector integrates the intensity profile from  $y=0$ to $y=y_{int}$ during the integration time period $T_{int}$. \label{fig:intensity_profile}}
\end{figure}

The beam radius of detection $R_d$ is defined as the maximum radial distance between the trajectory of the center of the scanning beam and the position of the receiving SC in the uncertainty plane, where the energy integrated by the detector on the receiving SC is sufficiently high to detect the beam with the required centroiding accuracy (assuming a matrix detector is used, see e.g.\cite{Gao2021laser}).  Note that this definition depends on the angular scan speed $\gamma$ and integration time $T_{int}$, and assumes that integration starts at the point of closest approach (worst case). This is shown in Fig. \ref{fig:intensity_profile}, where the receiving SC (white circle) accumulates energy along the red arrow from (x=$R_d$, y=0) to (x=$R_d$, $y_{int}$=$\gamma T_{int}$), while the laser beam moves along the spiral track indicated by the black arrow in vertical direction.
When the detector subsequently integrates in a second interval from $y_{int}$ to $2y_{int}$, the associated radius of detection $R_{d2}$ is smaller because the initial offset in y-direction requires a smaller lateral displacement in x-direction in order to accumulate the same amount of energy. 
The far-field beam intensity $I(x,y)$ is integrated over a path of length $\gamma T_{int}$ that is parallel to and laterally displaced from the spiral track by $R_d$. We shall assume that the far-field intensity is Gaussian distributed with waist $w$ and a maximum intensity that is given by $I_0=2P_0/(\pi w^2)$, where $P_0$ is the transmitted beam power. We then find from the underlying equations:
\begin{eqnarray}
E_{acc}&=& \frac{A_{ap}T_{l}}{\gamma}I_0 e^{\frac{-2 R_d^2}{w^2}}\int_0^{y_{int}}e^{\frac{-2 y^2}{w^2}}dt\nonumber\\
&=& \frac{A_{ap}T_{l}}{\gamma}\frac{P_0}{w\sqrt{2\pi}}e^{\frac{-2 R_d^2}{w^2}}\erf\left(\frac{\sqrt{2}y_{int}}{w}\right)\label{eq:accumulated_energy}
\end{eqnarray}
where $A_{ap}$ and $T_l$ are constants referring to the area of the entrance aperture and to the optical losses of the detection path, respectively.  
One can derive a similar expression that relates the minimum energy to the detection radius $R_d(n)$ in the integration interval from $(n-1) y_{int}$ to $n y_{int}$.
\begin{equation}
\begin{aligned}
E_{acc}&=&\frac{A_{ap}T_{l}}{\gamma}\frac{P_0}{w\sqrt{2\pi}}e^{\frac{-2 R_d(n)^2}{w^2}}\left[\erf\left(\frac{\sqrt{2}n y_{int}}{w}\right)-\right. \\
&&\left.\erf\left(\frac{\sqrt{2}(n-1) y_{int}}{w}\right)\right]\label{eq:accumulated segments}
\end{aligned}
\end{equation}
For a given critical energy that is required to detect the transmitted beam with sufficient SNR, we can use Eq.\ref{eq:accumulated_energy} to determine the radius of detection for the first and Eq. \ref{eq:accumulated segments} for any consecutive integration interval. This is shown on the right of Fig.\ref{fig:intensity_profile}, where we find that  $R_{d2}$=0.74 $R_d$ for the second integration interval, assuming $R_d$ is given by the beam waist.\\
Alternatively, Eq.\ref{eq:accumulated_energy} can be used to compute the radii of detection for different integration times, assuming they depend on the same critical energy threshold $E_{crit}$. Note that these relations are useful to analyze the results of Monte Carlo simulations and relate them to analytical predictions (see section \ref{sec:impact_high_freq_jitter}).\\
In the limit of very small integration time, $T_{int}\rightarrow 0$, we find that the radii of detection for successive integration windows approach the contours of a circle.
We refer to this limit as the ``Hard Sphere Model (HSM)'', as it defines a successful detection event whenever the distance between receiving SC and beam center is smaller than a fixed radius $R_{HS}$. The HSM has been used in previous studies of the acquisition process \cite{friederichs2016vibration, ma2021satellite}, where it can be applied if the received beam powers are high and the integration times short. This is the case for most missions dedicated to optical communication which are equipped with powerful lasers and large aperture telescopes. However, the HS model is unsuitable if the received optical power is low and long integration times are required, as is the case for gravitational wave observatories \cite{cirillo2009control} or missions without costly optical transceiver systems (see e.g.\cite{heinzel2017laser}).

\section{Correlations of the SC position between multiple tracks}
\subsection{Analytical model for detection on multiple tracks}
\begin{figure}
\centerline{\includegraphics[width=0.8\columnwidth]{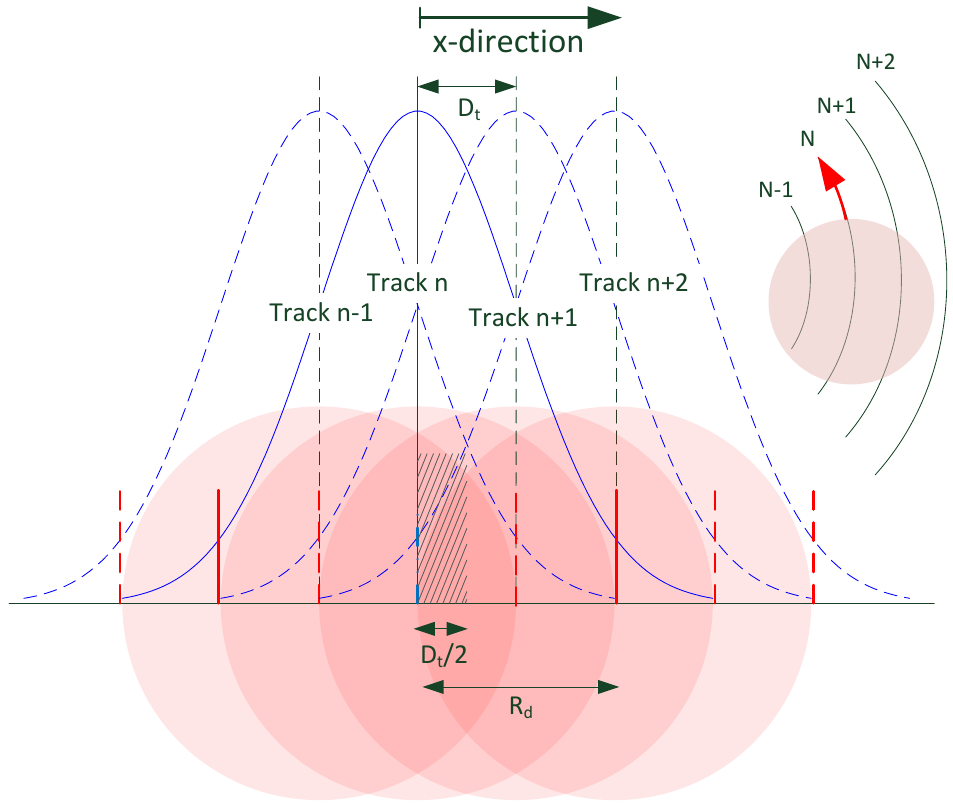}}
\caption  {Schematic of multiple overlapping tracks (n-1,n,n+1,n+2) during acquisition. \label{fig:overlapping_tracks}}
\end{figure}

\begin{figure}
\centerline{\includegraphics[width=0.8\columnwidth]{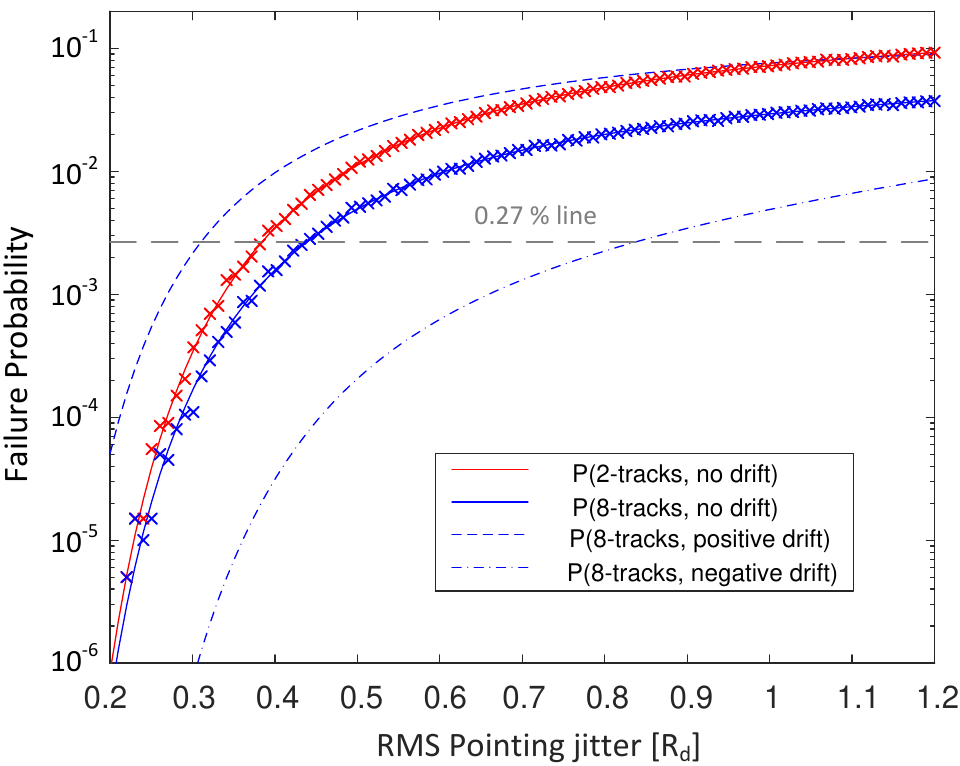}}
\caption  {Acquisition failure probability is plotted against the RMS level of beam jitter for an assumed track width of $D_t=0.8\,R_d$. The red and blue solid curves consider 2 and 8 tracks in the computation, respectively. The lower and upper dashed curves consider radially inbound and outbound drifts of the SC, respectively, with a magnitude of $R_d/4$ per revolution.\label{fig:multi-track_and_drifts}}
\end{figure}

The model in \cite{hechenblaikner2021analysis} derives the probability of acquisition failure for the case that a jittering scanning beam moves along the track of an Archimedean spiral in the uncertainty plane. 
The derivation of that model made several simplifying approximations. One of them was to consider only beam jitter in radial direction which is much more significant than jitter along track and fully accounts for the failure probability below a critical frequency, as discussed in section 4\ref{sec:jitter_during_passage}. 
Another approximation was to neglect correlations of the beam jitter between adjacent tracks and between individual integration windows which we investigate thoroughly in sections 4\ref{sec:jitter_between_tracks} and \ref{sec:MonteCarlo}, respectively.\\
The model also restricted beam detection to the two tracks closest to the SC in the uncertainty plane, assuming there is only small overlap of the beam profile between two adjacent spiral tracks and the jitter powers are low. 
In this section, we extend the previous model with the ability to detect the beam on multiple overlapping tracks, which allows studying the impact of large jitters and deterministic beam drifts on the detection probability. The overlap between tracks is defined by $OV=2 R_d-D_t$, where $D_t$ is the spiral pitch. If the pitch is made smaller and smaller, an increasing number of beam profiles overlap within the same region, as indicated in Fig.\ref{fig:overlapping_tracks}, where $D_t=R_d/2$ and tracks $n-1,n,n+1,n+2$ overlap. Without loss of generality (all other cases are covered by analogy) we make the assumption that the SC is located in between tracks n and n+1, but closer to track n (shaded area), and its location is given by the random variable $X_t$. The figure displays a cross-cut through the spiral plane along a radial vector from the spiral origin through the location of the SC. The beam intensity profiles are represented by the pink-shaded areas which all overlap in the black hatched area. The blue curves indicate the Gaussian intensity profiles along the radial direction and the red vertical lines indicate the radius of detection $R_d$, which in this example is located at the beam waist. For the mathematical derivation of failure probabilities we follow a similar approach as in \cite{hechenblaikner2021analysis}. 
The jitter-induced displacement of the laser beam while it moves past the location of the SC on track n can be equivalently modeled as a displacement of the SC (with opposite sign). We assume the latter to be represented by the random variable $X_n$, with analogous notations of the random variables for the other tracks.
The ``failure events'' $A_{n}$ and $B_{n}$ are defined to occur when the beam moves along track n and misses the SC on its left (smaller radius) or on its right (larger radius), respectively. Based on these definitions, the probabilistic inequalities listed in Equation \ref{eq:multi_track} can be derived. The individual rows define events $A$ and $B$ for tracks n-1, n, n+1, and n+2, respectively. 
\begin{align}
X_t+X_{n-1}&<-R_d-D_t & X_t+X_{n-1}&>R_d-D_t \nonumber\\
X_t+X_n&<-R_d & X_t+X_n&>R_d\nonumber\\
X_t+X_{n+1}&<-R_d+D_t & X_t+X_{n+1}&>R_d+D_t\nonumber\\
X_t+X_{n+2}&<-R_d+2D_t & X_t+X_{n+2}&>R_d+2D_t\label{eq:multi_track}
\end{align}
It is apparent that the random variable $X_t$ for the SC position is shared by all equations. In order solve these equations and derive an expression for the overall failure probability, it is useful to marginalize the variable $X_t$ and compute the conditional probabilities $P(A_n|X_t=x_t)$, $P(B_n|X_t=x_t)$ for events $A_n$ and $B_n$ to occur, given that the random variable $X_t$ assumes a specific value $X_t=x_t$.
The total probability $P_{n}$ that the receiving SC is missed while the beam moves along track $n$ is then given by the sum of the probabilities that it is missed on either the left or the right side of track n:
\begin{displaymath}
P_n(x_t)=P(A_n|X_t=x_t)+P(B_n|X_t=x_t),
\end{displaymath}
and similarly for the other tracks.
Assuming that the SC jitter is described by an unbiased normal distribution with variance $\sigma^2$ we find from Equations \ref{eq:multi_track} after integration over the SC jitter:
\begin{eqnarray}
P_{n-1}(x_t)&=&\frac{1}{2}\erfc\left(\frac{x_t+R_d+D_t}{\sqrt{2}\sigma}\right)+\frac{1}{2}\erfc\left(\frac{R_d-D_t-x_t}{\sqrt{2}\sigma}\right)\nonumber\\
P_{n}(x_t)&=&\frac{1}{2}\erfc\left(\frac{x_t+R_d}{\sqrt{2}\sigma}\right)+\frac{1}{2}\erfc\left(\frac{R_d-x_t}{\sqrt{2}\sigma}\right)\nonumber\\
P_{n+1}(x_t)&=&\frac{1}{2}\erfc\left(\frac{x_t+R_d-D_t}{\sqrt{2}\sigma}\right)+\frac{1}{2}\erfc\left(\frac{R_d+D_t-x_t}{\sqrt{2}\sigma}\right)\nonumber\\
P_{n+2}(x_t)&=&\frac{1}{2}\erfc\left(\frac{x_t+R_d-2D_t}{\sqrt{2}\sigma}\right)+\frac{1}{2}\erfc\left(\frac{R_d+2D_t-x_t}{\sqrt{2}\sigma}\right)\nonumber\\
&&\label{eq:multi_probability}
\end{eqnarray}
Assuming that the random variables for the jitter $X_{n-1}, X_n, X_{n+1}, X_{n+2}$ are uncorrelated between tracks, the combined probability of missing the SC for a given position $x_t$, $P_{tot}(x_t)$, can then be written as the product of the failure probabilities of the individual tracks. The general expression for failing to detect on $2N$ tracks labeled $n-N+1, ..., n+N$ is given by 
\begin{equation}
\begin{aligned}
P_{tot}(x_t)&=&\prod_{k=-N+1}^{N}P_n(x_t)=\prod_{k=-N+1}^{N}\frac{1}{2}\left[\erfc\left(\frac{x_t+R_d-k\cdot D_t}{\sqrt{2}\sigma}\right)\right.\\
&&\left.+\erfc\left(\frac{R_d+k\cdot D_t-x_t}{\sqrt{2}\sigma}\right)\right]\label{eq:P_tot_xt}
\end{aligned}
\end{equation}
We assume the density function of the marginal variable to be uniform and equal to $f(x_t)=2/D_t$ within the integration range from $0$ to $D_t/2$. 
When we weight $P_{tot}(x_t)$ with $f(x_t)$ and integrate over $x_t$, we obtain the total probability:
\begin{equation}
\label{eq:P_tot}
P_{tot}=\frac{2}{D_t}\int_{0}^{D_t/2} dx_t P_{tot}(x_t)
\end{equation}
The number of terms $2N$ which must be considered in the product depends on how strongly the tracks overlap and what the assumed level of beam jitter is. For track widths $D_t<R_d$ a total of more than 4-tracks is typically required for accurate results. This is demonstrated in Fig. \ref{fig:multi-track_and_drifts} where the red and blue solid lines denote the failure probabilities if 2 and 8 tracks are used in the computation, respectively. The two curves deviate from another with increasing beam jitter. 

\subsection{Reference mission parameters}
Table \ref{tab:reference_missions} introduces acquisition parameters for two virtual missions, which we will use as examples in the discussions of the following sections. The parameter $\sigma_{uc}$ refers to the width of an unbiased normal distribution defining the SC location in the uncertainty plane. The first row lists parameters for an earth observation mission on a low-earth orbit (mission scenario M1) which are derived from parameters for the GRACE FO mission \cite{heinzel2017laser} and were used previously in \cite{hechenblaikner2021analysis}. In this case, the received beam power is very low (~20 pW) due to a weak laser and a large beam divergence angle. After further attenuation by the receiver optics, only a few pW arrive at the detector, which requires long integration times and slow angular scan speed. This is similar to the case of gravitational wave observatories \cite{danzmann2000lisa}, where the very large distance between SC leads to similarly weak powers, although the divergence angle is about 50 times smaller, owing to the much larger telescope size. 
The parameters of mission scenario (M2) were mostly extracted from \cite{hindman2004} and correspond to a typical optical communication terminal in space, such as discussed in \cite{hindman2004,sterr2011beaconless,friederichs2016vibration}, where received powers are between 3 and 6 orders of magnitude higher and integration times correspondingly much shorter.
\begin{table}[h]
\centering
\caption{\label{tab:reference_missions}
Acquisition parameters for two virtual mission scenarios: (M1) Earth observation mission on low-earth orbit, and (M2) Laser communication link between SC.}
\begin{tabular}{ccccccc}
\hline
\textrm{}& \textrm{$R_d$}& \textrm{$D_t$}& \textrm{$\gamma$}& \textrm{$T_{int}$} & \textrm{$f_{crit}$} & \textrm{$\sigma_{uc}$} \\
\textrm{}& \textrm{$[\mu rad]$}& \textrm{$[\mu rad]$}& \textrm{$[\mu rad/s]$}& \textrm{$[s]$} & \textrm{$[Hz]$} & \textrm{$[\mu rad]$}\\
\hline
M1 & 135 & 203 & 67.5 & 1 & 0.125 & 872\\
M2  & 25   & 48 & 1.4E4 & 40E-6 & 140 & 291\\
\end{tabular}
\end{table}

\subsection{Beam drift}
\begin{figure}
\centerline{\includegraphics[width=0.8\columnwidth]{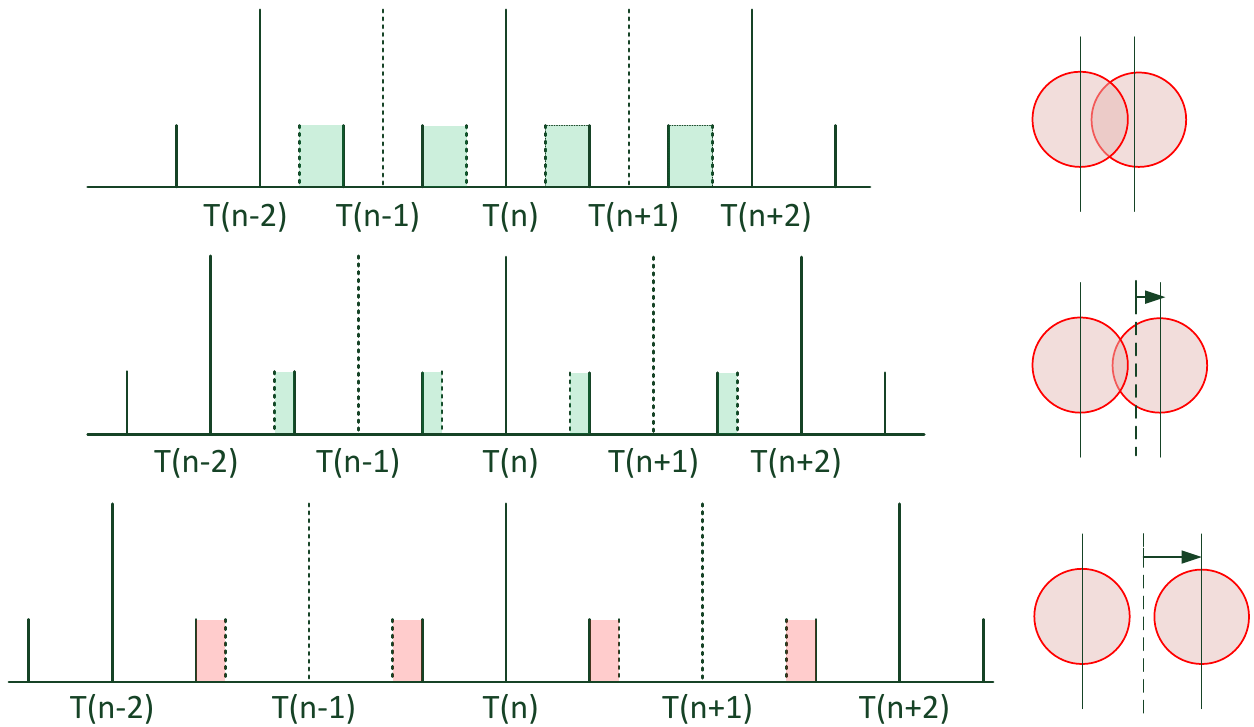}}
\caption  {The beam intensity profiles are plotted for adjacent tracks along a cross-cut in radial direction. The peak intensity and detection radii are represented by a central line and a pair of side-lines, respectively. The beam overlap (green shaded area) decreases from the top to the middle row and becomes negative, i.e. there is a gap between beams, in the bottom row (red shaded area)\label{fig:overlap_with_drifts}}
\end{figure}

It is important to note that the validity of the above derivation depends on the assumption that the position of the receiving SC remains fixed in the uncertainty plane of the scanning SC during the acquisition process. In reality, the scanning and receiving SC move in orbit and change their relative positions and attitudes. This is generally accounted for by the tracking system of the scanning SC which receives updated ephemeris of both SC before starting acquisition and estimates the required change of pointing due to their orbital evolution accordingly. Knowledge errors in the initial conditions, such as the relative SC velocity vectors, lead to drifts of the receiving SC in the uncertainty plane seen by the scanning SC\cite{sterr2011beaconless}.
For example, a radial velocity knowledge error of $v_r$ leads to a displacement of the SC in the uncertainty plane of $d=v_r\cdot 2\pi R_0/\gamma$ during one spiral revolution at a radial distance $R_0$. As the displacement error accumulates for each revolution, the SC performs a drift in the uncertainty plane. In a similar way, thermo-elastic drifts of the telescope assembly or the steering mirror impact the acquisition probability and must be budgeted accordingly. 
A deterministic beam drift with a constant displacement for each successive revolution leads to an effective increase (beam moves radially outwards) or decrease (beam moves radially inwards) of the track width $D_t$. By reciprocity, drifts of the SC position are equivalent to drifts of the beam in the opposite direction.
The effective track width $D_t'$ due to the drift speed $v_r$ then becomes 
\begin{equation}
D_t'=D_t+v_r 2\pi R_0/\gamma,
\end{equation}
where the velocity $v_r$ is positive for the beam moving radially outwards. The effect of a drift is shown in Fig. \ref{fig:overlap_with_drifts} which plots intensity peak and detection radii, indicated by the pair of lines surrounding it. The intensity lines are plotted for tracks $n-2$ to $n+2$ along a radial cross-cut through the spiral plane, where a different drift per track is shown in each row. The large spatial overlap for no drift in the top row (green area) shrinks due to a drift in the middle row (green area), and turns into a gap between beams for even larger drifts in the bottom row (red area). The upper and lower dashed blue curves of Figure \ref{fig:multi-track_and_drifts} indicate the failure probabilities for positive and negative drifts of a given magnitude, respectively. One can observe that drifts may have a significant impact on the failure probability and either increase or decrease it. 

In order to apply this to a practical example, we consider the acquisition parameters of a potential earth-observation mission which are given in the second row of Table \ref{tab:reference_missions} and aim to determine from Figure \ref{fig:multi-track_and_drifts} the maximum tolerable beam jitter and drift to guarantee a failure probability of less than 0.27\% ($3\sigma$).
The upper dashed blue curve yields a tolerable jitter RMS amplitude of $0.31\,R_d=42\,\mu rad$ for an assumed positive beam drift of $R_d/4$ per track. The latter corresponds to an uncertainty in the relative velocity of $v_r=R_d\gamma/(8\pi R_0)=0.24\,\mu rad/s$ under the conservative assumption that the remote SC is located in the outer-most track of the uncertainty region (at $R_0=3000\,\mu rad$). Assuming that the distance in between SC is $270\,km$,the relative velocity error must be smaller than $0.24\,\mu rad/s\times 270\,km=6.5\,cm/s$, which can easily be achieved by GPS-based precise orbit determination methods\cite{kang2003precise, hobbs2006precise}.

\section{Auto-correlation of the beam jitter}
\label{sec:jitter_correlation}
In the derivation of Eq.\ref{eq:P_tot_xt} we implicitly made two simplifying assumptions. One was that the jitter amplitudes are uncorrelated between tracks, i.e. $\langle X_n, X_{n+1}\rangle=0$, which allowed us to treat the associated random variables as independent. The other assumption was that the beam jitter is fully correlated while the beam sweeps past the SC, which allowed us using a single random variable to represent the jitter amplitude during this period. These two assumptions are investigated in more detail in this section, where we will observe that the properties of the auto-correlation function strongly impact the failure probability.

\subsection{Jitter correlations between tracks}
\label{sec:jitter_between_tracks}
\begin{figure}
\centerline{\includegraphics[width=0.8\columnwidth]{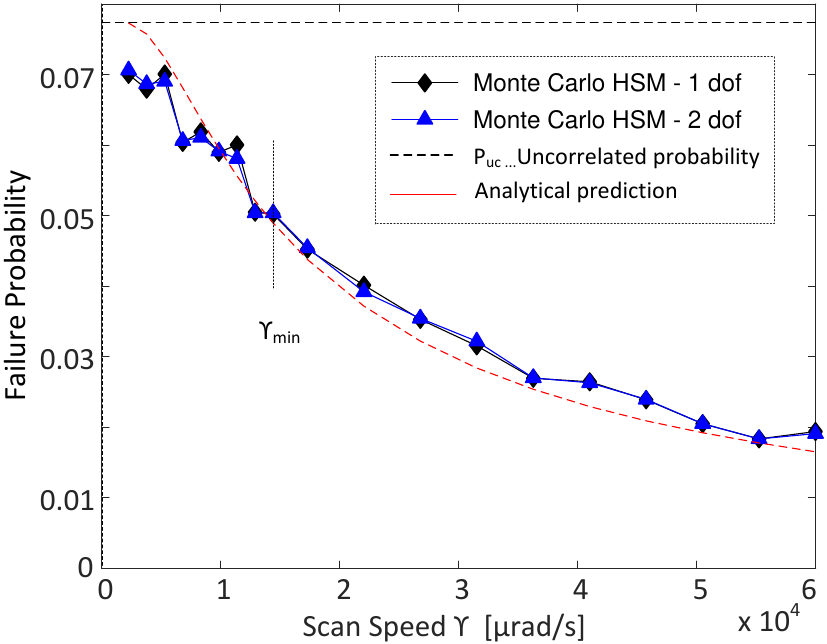}}
\caption  {The failure probability is plotted against the scan speed for the results of Monte Carlo simulations with 1d (black diamonds) and 2d (blue triangles) jitter. The analytical prediction and the uncorrelated limit are given by the red and black dashed lines, respectively.\label{fig:failure_against_gamma}}
\end{figure}

Assuming that the jitter amplitude is uncorrelated between tracks, i.e. $\langle X_n, X_{n+1}\rangle=0$, the total failure probability is computed as the product of the individual probabilities to miss the SC on the respective track (see Eq. \ref{eq:P_tot_xt}). This is only valid if the auto-correlation function is much smaller than the mean time of one spiral revolution which is given by 
\begin{equation}
T_{rev}=\sqrt{\pi/2}\,\sigma_{uc}2\pi/\gamma.
\label{eq:mean_time_between_tracks}
\end{equation}
Using the parameters of Table \ref{tab:reference_missions} we find that $T_{rev}$=102\,s for scenario M1 and $T_{rev}$=0.164\,s for scenario M2. 
The auto-correlation function $r_{xx}(\tau)$ can be obtained from the 2-sided power spectral density $S(f)$ through the Wiener-Khinchin theorem:
\begin{equation}
\label{Eq:Wiener}
r_{xx}(\tau)=\langle x(t) x(t+\tau)\rangle=\int^{\infty}_{-\infty}df S(f) e^{2\pi i\tau f}
\end{equation}
Based on an available jitter model for the SILEX mission,  the analysis in \cite{teng2018optimization} uses a colored noise spectrum with constant spectral density up to a roll-off frequency $f_{r}$=1\,Hz,  from where it drops as $1/f^2$ towards higher frequencies: $S(f)=160 \mu rad^2/Hz/(1+f/f_{r})^2$. 
We can calculate $r_{xx}$ for this spectrum using Eq.\ref{Eq:Wiener} and find that 
\begin{equation}
r_{xx}(\tau)=e^{-2\pi|\tau|f_r}. 
\label{eq:colored_noise_correlation}
\end{equation}
Evaluating $r_{xx}$ for the revolution times given above we find that the correlation is completely negligible for $T_{rev}$=102\,s and has dropped to 36\% of its peak value for $T_{rev}$=0.164\,s. Therefore, even for configuration M2 correlations are still small enough to fall within the applicability range of Eq.\ref{eq:P_tot}.\\ 
However, if the scan speed  $\gamma$ is increased even more, this also increases the correlation between tracks. For example, increasing $\gamma$ in scenario M2 by a factor 4 we find that $r_{xx}$ increases to 77\%. 
The effect of such positive correlations is to decrease the failure probability because the beam is offset by the same distance from the nominal path on two adjacent tracks, which ensures that it is detected on one track if it was missed on the other. In the extreme case of fully correlated variables, this can be easily proven mathematically with the help of Eqs.\ref{eq:multi_track}.
Using Eqs.\ref{eq:colored_noise_correlation}, \ref{eq:mean_time_between_tracks} and requiring $r_{xx}>1/e$, we derive Eq.\ref{eq:scan_speed_limit} for the minimum scan speed in order to have strong jitter correlations between tracks that reduce the failure probability. Note that by inversion of Eq.\ref{eq:scan_speed_limit} we can also obtain an expression for the maximum roll-frequency where strong correlations occur for a given scan speed.
\begin{equation}
\gamma_{min}=f_r\sqrt{\pi/2}(2\pi)^2\sigma_{uc}
\label{eq:scan_speed_limit}
\end{equation}
In order to investigate the dependency of failure probability on scan speed, we performed Monte Carlo simulations using a ``Hard-Sphere'' detection model and the parameters of Table \ref{tab:reference_missions} for scenario M2. The results are given in Fig.\ref{fig:failure_against_gamma} by the black and blue lines for 1d (radial) and 2d(radial and axial) jitter, respectively. Assuming that the probability of a hit increases as the auto-correlation function evaluated for the mean revolution time, $r_{xx}(T_{rev})$, and assuming it rises from the ``uncorrelated'' limit $P_{uc}$ given by Eq.\ref{eq:P_tot}, we obtain for the failure probability: $P_{fail}=P_{uc}[1-r_{xx}(T_{rev})]$. This curve is plotted as the red-dashed line in Fig.\ref{fig:failure_against_gamma}. It is in fair agreement with the Monte Carlo results, suggesting the validity of the approximations we made above.
It should be noted that the authors of \cite{hindman2004} have found before from a Monte Carlo simulation that the probability of success decreases with increasing scan time (i.e. decreasing scan speed), but they made no further investigations into the origin of this effect.\\
In summary, the acquisition architect should consider in the system design that it is favorable to increase scan speeds to above $\gamma_{min}$ as this leads to a significant reduction of failure probability compared to the uncorrelated limit of Eq.\ref{eq:P_tot}.
\subsection{Jitter correlations during beam passage}
\label{sec:jitter_during_passage}
\begin{figure}
\centerline{\includegraphics[width=0.8\columnwidth]{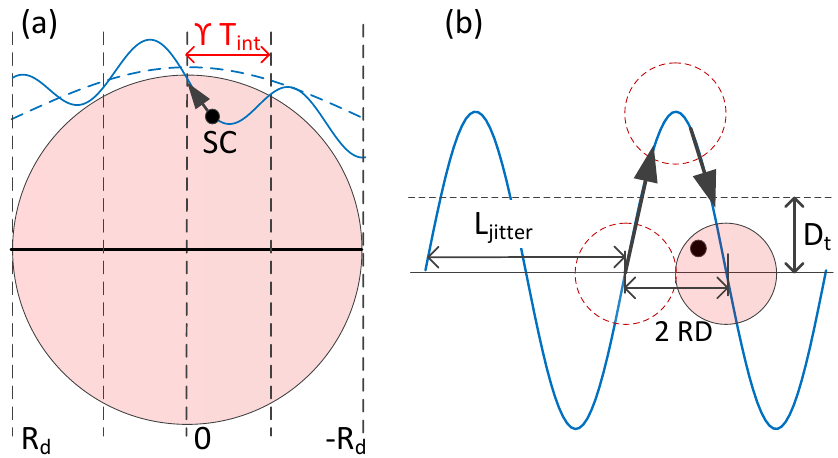}}
\caption  {(a) Detection by the SC in one of 4 integration windows while the beam passes by. (b) For beam oscillations frequencies of $f_{crit}$ or higher, the SC is always detected.\label{fig:jitter_in_windows}}
\end{figure}

In this section we perform a qualitative analysis of the effects of higher jitter frequency on the probability of detection. The arguments and derivations will be verified by comparison to Monte Carlo simulations in section \ref{sec:MonteCarlo}.
Two important time-scales affect the detection of the scanning beam:\\ 
\emph{(1) Detector integration time $T_{int}$}: The detector integrates during $T_{int}$ and approximately averages out fluctuations that occur at frequencies higher than the Nyquist frequency $f_{ny}$=1/(2$T_{int}$). Assuming band-limited white Gaussian noise (BLWGN), successive integration windows become uncorrelated, if the jitter spectrum extends up to $f_{ny}$, which follows from the sampling theorem.\\ 
\emph{(2) Beam passage time $T_{pass}$}: The beam passage time is defined as the time it takes for the beam to sweep past the receiving SC in the uncertainty plane. It is given by $T_{pass}$=2$R_d/\gamma$.\\
Therefore, if $T_{int}\leq T_{pass}$, the integration windows become uncorrelated for frequency spectra that extend beyond a fundamental critical frequency $f_{crit}$:
\begin{equation}
f_{crit}=\frac{\gamma}{4R_d}
\label{eq:f_crit}
\end{equation}
In this case, the total failure probability $P_{tot,all}$ for all windows is simply given by the product of the individual probabilities $P_{tot}(w_i)$ for each integration window 'i', which generally leads to a strong decrease of the total failure probability in comparison to the one of a single window:
\begin{equation}
P_{tot,all}=P_{tot}(w_1)\times P_{tot}(w_2)..\times P_{tot}(w_n)=P_{tot}^n(w_1)
\end{equation}
For the case of correlated integration windows, we have to use conditional probabilities to obtain the total probability
\begin{eqnarray}
P_{tot,all}&=&P_{tot}(w_1)\times P_{tot}(w_2|w_1)..\times P_{tot}(w_n|w_{n-1},.., w_1)\nonumber\\
&=& P_{tot}(w_1),
\end{eqnarray}
where the expression on the second line is obtained for full correlation. Figure \ref{fig:jitter_in_windows}(a) shows the individual integration windows during beam passage. Low frequency jitter (dashed blue line) is not detected in any of the windows, while the addition of high-frequency jitter (solid blue line) leads to detection in one of the windows. Note that in this representation we imposed the jitter on the SC and left the beam static, which is equivalent to the actual case (the other way round). \\ 
The critical frequency $f_{crit}$ can also be derived considering that the width $\tau_{wn}$ of the auto-correlation function of BLWGN is given by $\tau_{wn}$=$1/(2f_{cut})$, where $f_{cut}$ is the cut-off frequency. The failure probability decreases if the spatial width of the auto-correlation $\gamma\tau_{wn}$ is smaller than the beam diameter (2$R_d$). From this relation one finds again the critical frequency of Eq.\ref{eq:f_crit}.\\
Another intuitive picture is given in Figure \ref{fig:jitter_in_windows}(b). A fluctuation of frequency $f$ is plotted in the spatial domain where this corresponds to a wavelength of $\lambda$=$\gamma/f$. We notice that if half the wavelength equals $2R_d$, the fluctuating beam, which moves along the blue trajectory, will always sweep across the area where the SC (indicated by a black circle) is located. If the wavelength becomes larger, this is not the case. This relationship yields again the critical frequency. 

We conclude that frequencies in the jitter spectrum above $f_{crit}$ reduce the failure probability, as the individual integration windows become uncorrelated. In this case, the analytical model of Eq.\ref{eq:P_tot} provides a worst case estimate $P_{tot}$ of the actual failure probability, if we use in the model as noise amplitude ($\sigma$) a value derived from the integral over the jitter power up to the critical frequency.
\begin{equation}
\sigma=\sqrt{2\int_0^{f_{crit}}PSD(f) df}
\end{equation}
Note that this relationship only holds as long as the RMS amplitude of the total jitter power, integrated over the full spectrum, does not exceed a value of $\approx 2R_d$. Larger fluctuations are likely to displace the beam by so much that the SC is outside of its detection range on either side of the track.

\section{Monte Carlo simulations for slow scans}
\label{sec:MonteCarlo}
In this section we aim to verify the discussion of section 4\ref{sec:jitter_during_passage} and analyze the failure probability when varying the jitter spectrum in a way that affects the correlations between integration windows during beam passage. This applies to the limit of slow scans with long integration times, resulting in a low critical frequency. We therefore baseline mission scenario M1 of Table 1 and investigate the failure probability for spectral cut-off frequencies below $f_{crit}$  (subsection 5\ref{sec:low frequency jitter}) and above $f_{crit}$ (subsections 5\ref{sec:impact_high_freq_jitter} and 5\ref{sec:very_high_jitter}). 
\subsection{Simulation approach}
\label{sec:simulation_approach}
\textbf{Generating the noisy scan trajectory}\\
In our Monte Carlo (MC) simulations the nominal spiral track is perturbed by injection of noise, which is generated according to a given noise spectrum, into the propagator of the scanning beam. 
In order to generate a time series with the required noise spectrum, we imprint a random phase on the linear spectral density and select a random amplitude for each frequency bin. The amplitude is generated from a Gaussian distribution of width $\sigma_{amp}$ which corresponds to the amplitude of the spectral density at the given frequency. Following the approach of \cite{timmer1995generating}, we then perform the Inverse Fourier Transform (IFFT) to obtain a random time series.\\
For some simulations, the time series is then superposed on the unperturbed spiral track  in radial direction only (1 dof), which represents the case where the scanning beam is steered by the body-pointed SC, as is done for example in gravitational wave observatories\cite{cirillo2009control}. For other simulations, we generate a second time series and additionally superpose it in tangential direction (2 dof), which represents the case where a beam-steering mirror is used on a jittering SC platform\cite{hindman2004, friederichs2016vibration}. Note that in this case the total jitter amplitude is the root-sum-squared of the two degrees of freedom  and therefore $\sqrt{2}$ higher than for radial jitter alone.\\
For each simulation run, a random position of the SC in the uncertainty plane is selected through coordinates ($\alpha$, $\beta$) which are generated from a Gaussian distribution of width $\sigma_{uc}$. When the propagator moves the scanning beam along the perturbed spiral track, the SC may detect the beam on a given spiral track by one of three selectable methods which are described below.\\*[0.5 ex]
\textbf{Detection Models}\\
\emph{(1) The Hard Sphere Model (HSM)} detects the beam if its distance to the SC is closer than $R_d$ at any time\cite{friederichs2016vibration}. This model can only be applied if received beam powers are high and integration times short (see section \ref{sec:R_d}).\\
\emph{(2) The Linear integration model (LIM)} detects the beam if its radial distance to the SC, averaged within the given integration time $T_{int}$,  is smaller than $R_d$. Additionally, integration windows are constrained to be aligned in time with the point of closest approach between SC and nominal spiral track, which corresponds to a worst case scenario that is considered in our definition of $R_d$. However, during the integration period the beam may fluctuate around its mean displacement so that the detector accumulates a certain amount of energy for time periods where the beam is either displaced farther away or closer to the SC with respect to its mean position. Taking only the mean position into account in the LIM amounts to a “linear intensity approximation”, where energy gain for the beam closer to the SC by a distance $\Delta_x$ is considered to balance the energy loss for the beam farther away by a distance –$\Delta_x$. This simplification allows separating the impact of window correlation from other effects and therefore supports a thorough understanding of the simulation results.\\
\emph{(3) The Non-Linear Physical Model (NLPM)} detects the beam if the integrated energy exceeds a critical threshold energy during any integration period. This model is representative of the physical reality, where the nonlinear  Gaussian beam profile moves along the perturbed spiral track and the SC detector accumulates energy continuously during each successive integration window. As the Gaussian beam intensity profile is non-linear and convex at a distance $R_d$ from its center, a SC located at this position integrates more power for a fluctuating than for a stationary beam. Owing to this non-linearity, the LIM (and likewise the analytical model) underestimate the detection probability compared to the exact predictions of the NLPM. In the NLPM the integration windows line up randomly with respect to the point of closest approach. This additionally increases the probability of detection with respect to the LIM, because the accumulated energy during an integration period is always larger if the point of closest approach lies within the integration window than if integration starts from there.\\*[0.5 ex]
\textbf{Comparing integration times}\\
Apart from the choice of detection model, we can also select either 1\,s or 2\,s integration time.  In order to compare between these two options, we can choose between 2 methods:\\
\emph{(1) Fixed $R_d$:} The radius of detection $R_d$ is set to be the same for both integration times. This implies that the critical energy threshold differs between the two options and is computed as the integrated energy over the interval $y_{int}$=$\gamma T_{int}$, respectively.\\
\emph{(2) Fixed $E_{crit}$:} The critical energy threshold is set to be the same for both integration times. In this case, a specific radius of detection is defined for $T_{int}$=2\,s for which the associated critical energy threshold $E_{crit}$ is calculated that is also used for $T_{int}$=1\,s. Using this threshold, the detection radii for the outer windows are computed, following the approach of section \ref{sec:R_d}.\\*[0.5 ex]
For each specific choice of simulation parameters, we performed a total of between 6E4 to 12E4 Monte Carlo runs.  In order to investigate the system behavior, we typically varied either the total power (i.e. the RMS amplitude) distributed over a selected spectral noise shape or we varied the spectral cut-off frequency while keeping the power spectral density constant.

\subsection{Low frequency jitter}
\label{sec:low frequency jitter}
\begin{figure}
\centerline{\includegraphics[width=0.8\columnwidth]{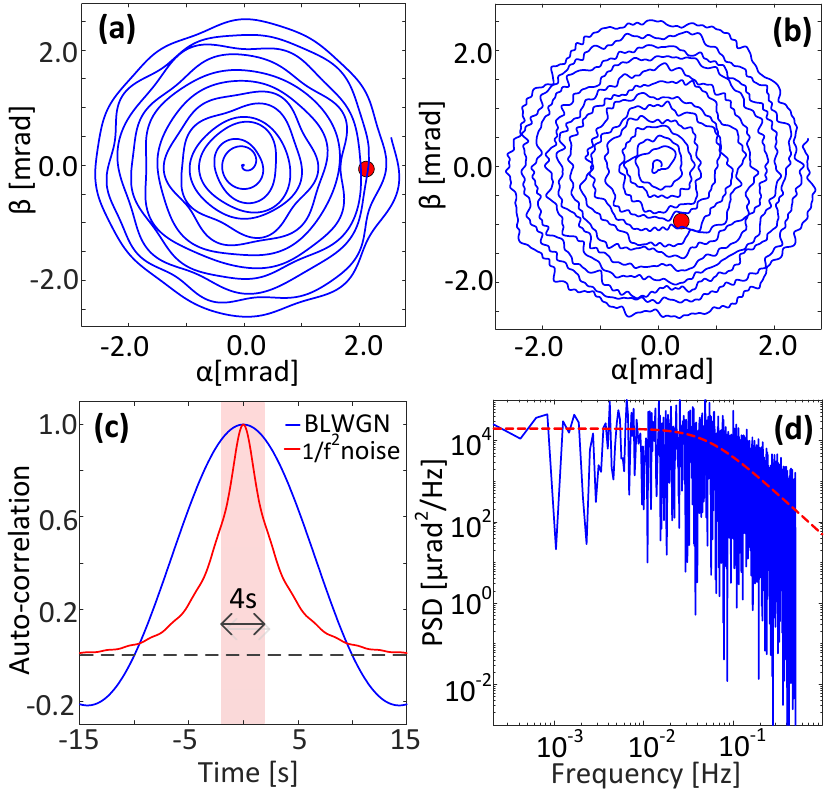}}
\caption  {Noisy scan pattern for (a) BLWGN and (b) $1/f^2$ noise. (c) Auto-correlation functions for BLWGN and $1/f^2$ noise are compared to $T_{pass}$=4\,s (red-shaded area). (d) PSD of scan pattern for $1/f^2$ noise.\label{fig:sample_trajectories}}
\end{figure}

\begin{figure}
\centerline{\includegraphics[width=0.8\columnwidth]{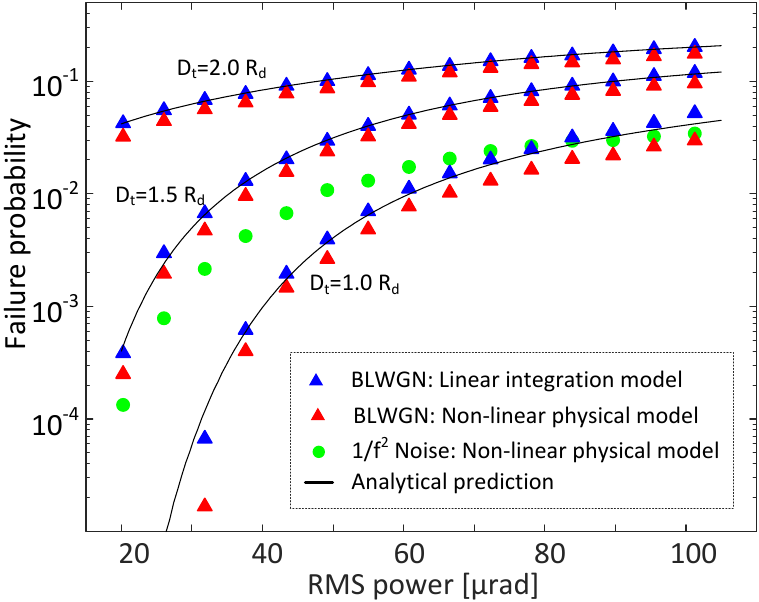}}
\caption  {Failure probabilities are plotted against RMS amplitude for track widths $D_t$=2.0, 1.5, and 1.0\,$R_d$ (top to bottom). Results for BLWGN from LIM and NLPM are given by blue and red triangles, respectively. Analytical predictions are indicated by the black lines, and results for $1/f^2$ noise using the NLPM by the green circles. \label{fig:MC_low_frequency}}
\end{figure}

In this section we look at the Monte Carlo (MC) analysis results for the failure probability due to beam jitter that is limited to a spectral range below the critical frequency. In this case, as discussed in section 4\ref{sec:jitter_during_passage}, the width of the auto-correlation function is larger than the time $T_{pass}$=4\,s it takes the beam to sweep past the SC in the uncertainty plane, as shown in Fig.\ref{fig:sample_trajectories}(c). Individual integration windows are fully correlated so that the probability of detecting the beam in one window is equal to the joint probability of detecting it in any of the windows.\\
We performed simulations for the example of band-limited white Gaussian noise that has a flat linear spectral density (LSD) from 0\,Hz up to a cut-off frequency $f_{cut}$=50\,mHz and for $1/f^2$ colored noise with a roll-off frequency of $f_r$=50 mHz, where we used the standard parameters for mission M1 in Table \ref{tab:reference_missions}. Examples for scan patterns are given in Fig.\ref{fig:sample_trajectories}(a) and (b), respectively. The colored noise PSD computed from the respective scan pattern is plotted in Fig.\ref{fig:sample_trajectories}(d) and agrees well with the theoretical curve (red-dashed line).\\
Figure \ref{fig:MC_low_frequency} plots the failure probability against the RMS jitter amplitude for various track widths and a given cut-off frequency $f_{cut}$=50\,mHz. The theoretical predictions from Eq.\ref{eq:P_tot} are given by the black lines, the Monte Carlo results for BLWGN obtained from the LIM and the NLPM with 1\,s integration time are given by the blue and red triangles, respectively. The LIM results agree very well with the analytical predictions while the NLPM yields slightly lower failure probabilities, as expected,  because the accumulated energy is higher for non-linear averaging. In contrast, the results for $1/f^2$ noise obtained from the NLPM for $D_t$=1.5\,$R_d$, indicated by the green circles, are significantly lower than the analytical predictions. The reason for this deviation is the power at frequencies above 50\,mHz. Consequently, the auto-correlation function width decreases to 60\% during beam passage (see Fig.\ref{fig:sample_trajectories}(c) and the integration windows become partially uncorrelated, thereby decreasing the total failure probability. We will investigate this in more detail in the next section.
 
\subsection{High frequency jitter and dimensionality}
\label{sec:impact_high_freq_jitter}
\begin{figure}
\centerline{\includegraphics[width=0.8\columnwidth]{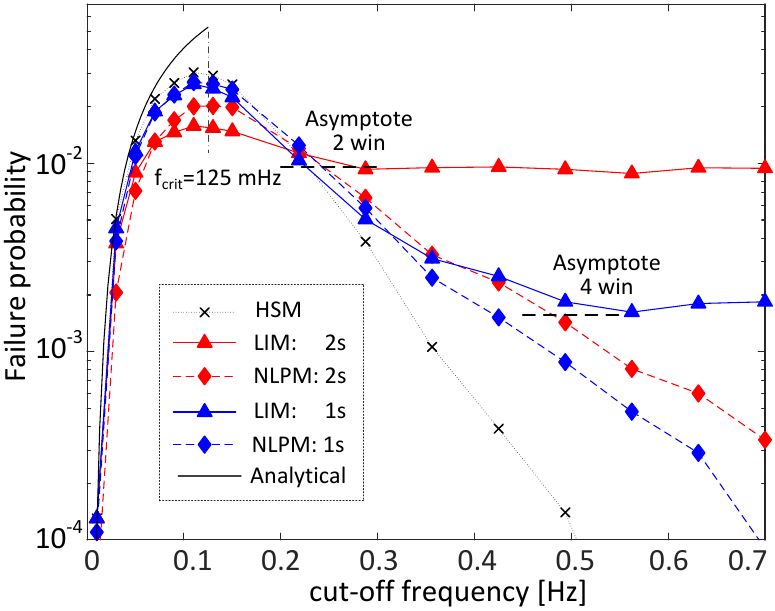}}
\caption  {Failure probability is plotted against spectral cut-off frequency for 1\,s (blue) and 2\,s (red) integration times, where LIM and NLPM-1dof predictions are represented by solid and dashed lines, respectively. HSM-1dof and analytical predictions are given by black-dotted and black-solid lines, respectively.\label{fig:MC_fixed_RD}}
\end{figure}

In this subsection we investigate the impact of high frequency jitter on the failure probability using the acquisition parameters of mission M1 in Table \ref{tab:reference_missions}. To this purpose, we use a BLWGN spectrum of fixed linear density $A_{lsd}=124\,\mu rad/\sqrt{Hz}$ and progressively increase the cut-off frequency. Consequently, power at higher frequencies is added to the spectrum at each step and the associated effect can be studied. Fig.\ref{fig:MC_fixed_RD} plots the failure probabilities for 1\,s (blue lines) and 2\,s (red lines) integration time as predicted by the LIM (solid lines) and NLPM (dashed lines). The Monte Carlo results are all close to the analytical prediction (black solid line) for low cut-off frequencies and peak at a cut-off frequency corresponding to $f_{crit}$, as expected from our discussion in section 4\ref{sec:jitter_during_passage}. 
The failure probabilities predicted by the LIM then decrease again and approach an asymptotic limit which agrees very well with an analytical prediction. The asymptotic limit is approximately reached when the cut-off frequency equals the Nyquist frequency $f_{ny}$=1/(2$T_{int}$) and the windows become fully uncorrelated. This occurs when $f_{cut}$=250\,mHz and $f_{cut}$=500\,mHz for 2\,s and 1\,s integration time, respectively.\\
The asymptotic limit can be computed analytically by multiplying the integration windows, which amounts to taking the square of Eq.\ref{eq:P_tot_xt} for $T_{int}$=2\,s (2 windows) and taking it to the fourth power for $T_{int}$=1\,s (4 windows) before integrating over $x_t$. Note for 4 windows the values for the detection radii of the two outer windows must be decreased slightly as discussed in section \ref{sec:R_d}, because the beam intensity is lower than for the 2 inner windows.\\
The NLPM results, plotted for jitter in radial direction only, follow fairly well those of the LIM for low frequencies and up to a point that is close to the asymptotic limit at the Nyquist frequency. After this point the NLPM failure probabilities continue decreasing while the LIM curve flattens out. This is because the non-linear beam intensity profile leads to an increase of accumulated energy with increasing beam jitter, thereby decreasing the failure probability, as discussed before in section 5\ref{sec:simulation_approach}. The results for the HSM, which correspond to the limit of an infinite number of integration windows, are given by the black-dotted line. Increasing the number of integration windows in the NLPM should therefore approach the HSM, which is what we observe by comparing the the curves for 1\,s (blue-dashed) and 2\,s (red-dashed) integration time to the one of the HSM, noting that the curve for 1\,s is much closer to it than the one for 2\,s. We also observe that the failure probability for 1\,s lies below the ones for 2\,s for frequencies above $f_{crit}$, meaning that it is more efficient to use shorter integration windows if $R_d$ remains constant. However, we found from other simulations that if the critical energy threshold is kept constant, which implies that $R_d$ is smaller for 1\,s than for 2\,s integration time, the opposite holds, i.e. it is more efficient to use longer integration times.

\begin{figure}
\centerline{\includegraphics[width=0.8\columnwidth]{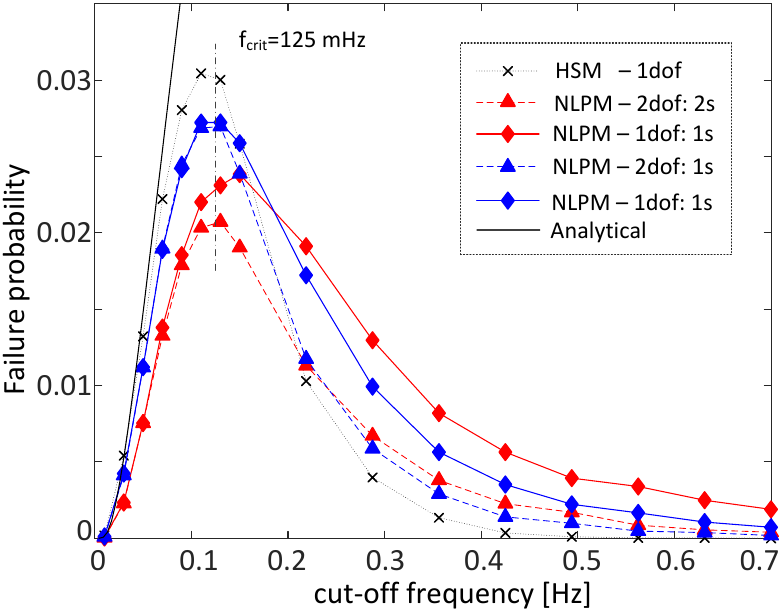}}
\caption  {Failure probability is plotted against spectral cut-off frequency for 1\,s (blue) and 2\,s (red) integration times. NLPM-1dof and NLPM-2dof predictions are represented by dashed and solid lines, respectively. HSM-1dof and analytical predictions are given by black-dotted and black-solid lines, respectively.\label{fig:MC_1dof_2dof}}
\end{figure}

We now investigate the difference in failure probability between 1-dimensional (radial) and 2-dimensional (radial and tangential) jitter, where the former corresponds to the trajectory of a beam pointed by the SC and the latter to the trajectory of a beam pointed by a stable steering mirror on a vibrating SC. The results are shown in Fig.\ref{fig:MC_1dof_2dof} for the same acquisition parameters as used for Fig.\ref{fig:MC_fixed_RD} but with a linear instead of a semi-logarithmic scaling. We observe that the data for 2-dimensional jitter agree exactly with those for 1-dimensional jitter up to the critical frequency $f_{crit}$, after which they start diverging but follow the same general trend. As expected, the curve for 2-dimensional jitter consistently lies above the one for 1 dimensional-jitter for higher cut-off frequencies, but the tangential component has not effect for cut-off frequencies below $f_{crit}$.

\subsection{The limit of very high jitter power}
\label{sec:very_high_jitter}
\begin{figure}
\centerline{\includegraphics[width=0.8\columnwidth]{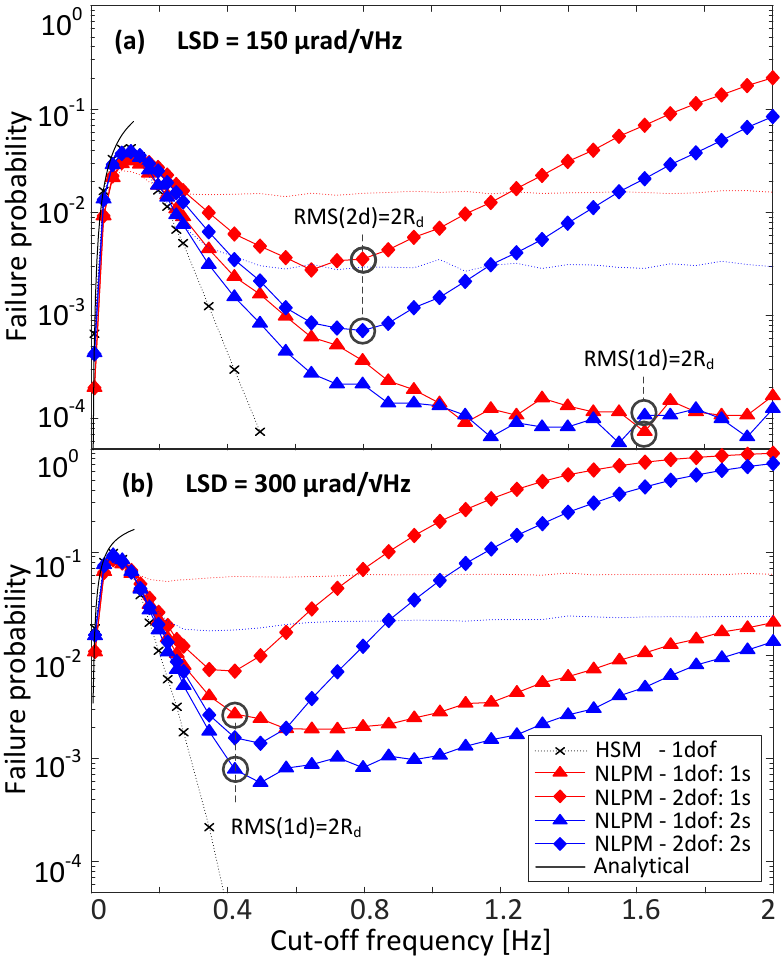}}
\caption  {Failure probability is plotted against spectral cut-off frequency for 1\,s (blue) and 2\,s (red) integration times; NLPM-1dof (triangles), NLPM-2dof (diamonds), HSM-1dof(black-dotted line), analytical predictions (black-solid line).\label{fig:MC_high_RMS}}
\end{figure}

In this subsection we investigate what happens if the RMS jitter amplitude is increased to a very high level that exceeds the diameter (2$R_d$) of the laser beam.
The results of MC simulations with the parameters given for mission M1 in Table \ref{tab:reference_missions} are shown in Fig.\ref{fig:MC_high_RMS}, where a BLWGN spectrum with a linear spectral density of either $150\,\mu rad/\sqrt{Hz}$ (a) or $300\,\mu rad/\sqrt{Hz}$ (b) and an extended cut-off frequency range between [0, 2.0] Hz was used.\\ 
In Fig. \ref{fig:MC_high_RMS}(a) we observe that the failure probabilities for 2d jitter (diamonds) decrease from a peak at $f_{crit}$ to a minimum around $f_{min}$=0.8\,Hz, where the RMS jitter amplitude corresponds to $2\,R_d$, before rising again monotonously.
The rise of failure probabilities for 1d jitter (triangles) is expected to occur around a frequency twice as high, because the linear spectral density is lower by $\sqrt{2}$ and therefore needs to be integrated over a larger range in order to obtain an RMS amplitude of $2\,R_d$. The respective minimum probability is very low and close to the edge of the plot so that the rise of probability is not well discernible.\\
In order to see this more clearly, we increase the linear spectral density of the noise by a factor 2 in Fig.\ref{fig:MC_high_RMS}(b) and observe that the failure probabilities decrease from their peak to a minimum close to the predicted frequency of $f_{min}$=0.4\,Hz, where the RMS jitter amplitude exactly equals the beam diameter. These observations verify our qualitative notion of section 4\ref{sec:jitter_during_passage} that increasing the amplitude of jitter beyond a level corresponding to the beam diameter leads to an increasing loss of accumulated energy and an associated rise of failure probability.\\
The predictions of the LIM, given by the dotted red and blue lines in (a) and (b), are found to settle at the asymptotic value we discussed above, which reflects the fact that it does not account for the non-linearity in the beam intensity profile when integrating the received power during $T_{int}$. 
Similarly, the results for the HSM (black dotted line) do not predict the physical reality, as the HSM detects a ``hit'' of a wildly fluctuation beam that enters a region delimited by $R_d$ around the SC, even if the time spent within this region is infinitely short and the accumulated energy consequently zero. For this reason, the failure probabilities predicted by the HSM continue decreasing while the ``true'' predictions of the NLPM rise for jitter amplitudes that exceed $2\,R_d$ and converge to a 100\% failure probability for very large amplitudes.

\section{Conclusion}
In summary, in this paper we present a comprehensive investigation on the impact of correlations of laser beam jitter on the acquisition probability of optical links in space.
We first derived an analytical model that considers correlations between the SC position and multiple overlapping scanning tracks, enabling us to assess the effect of deterministic beam drifts in the uncertainty plane, and derived an expression for the maximum tolerable drift rate.\\
We then investigated the limit of high scan speeds and short integration times, where positive correlations of beam jitter between adjacent tracks leads to a decrease of the failure probability, and derived an expression for the minimum scan speed necessary to exploit this effect in the system design. In the opposite limit of slow scan speeds $\gamma$ and long integration times, we determined a critical frequency $f_{crit}=\gamma/(4R_d)$ for spectral cut-off of the beam-jitter,  above which the integration windows become increasingly uncorrelated. We showed how this effect, coupled with the effect of non-linearities when the detector integrates across the convex beam profile, strongly decrease the failure probability for increasing jitter frequencies. Therefore, adding noise to the jitter spectrum above the critical frequency may improve the acquisition performance up to several orders of magnitude, as long as the total jitter RMS does not exceed the beam diameter, which can also be exploited in the system design.
As far as spectral properties of the beam jitter are concerned, the major derivations in this paper depend only on the width of the auto-correlation function $\tau_{corr}$ which is compared to the mean time between tracks in section 4\ref{sec:jitter_between_tracks} and to the beam passage time in section 4\ref{sec:jitter_during_passage}. Therefore, our results can be generalized to other spectral shapes than band-limited white noise and $1/f^2$ noise when considering the respective expression for $\tau_{corr}$.\\ 
Our analytical models and derivations were verified against the results of Monte Carlo simulations which have been performed for a variety of different detection models, for 1d (radial) and 2d (radial and tangential) jitter, for band-limited white Gaussian and colored noise, and for two different operational regimes.
Our analyses support a comprehensive understanding of the impact of jitter spectra and correlations on acquisition probability and facilitate the optimal choice of system parameters for a broad range of mission scenarios, from optical communication terminals to interferometric science missions in space. 
\begin{backmatter}
\bmsection{Acknowledgments} The author gratefully acknowledges fruitful discussions with Tobias Lamour, Simon Delchambre, Tobias Ziegler, Christian Greve, and R\"udiger Gerndt (Airbus).

\bmsection{Disclosures} The author declares no conflicts of interest.

\bmsection{Data availability} The data underlying the results presented in this paper were generated using analytical formulas presented in the text as well as Monte Carlo simulations. Simulation parameters and data can be provided upon reasonable request.

\end{backmatter}


\end{document}